\documentclass[letters,usenatbib]{mnras}

\usepackage[T1]{fontenc}

\DeclareRobustCommand{\VAN}[3]{#2}
\let\VANthebibliography\thebibliography
\def\thebibliography{\DeclareRobustCommand{\VAN}[3]{##3}\VANthebibliography}

\usepackage{graphicx}	
\usepackage{amsmath}	
\usepackage{amssymb}	
\usepackage{newtxtext,newtxmath}
\usepackage{easyReview}

\title[Helium abundance in SIRs]{New insights on the behaviour of solar wind protons and alphas in the Stream Interaction Region in solar cycle 23 and 24  }

\author[Yogesh et al.]{
Yogesh,$^{1,2}$\thanks{E-mail: yphy22@gmail.com}
D. Chakrabarty$^1$,
and Nandita Srivastava$^3$
\\
$^1$Physical Research Laboratory, Navrangpura, Ahmedabad 380009, India\\
$^{2}$Indian Institute of Technology-Gandhinagar, Gandhinagar 382055, India\\
$^{3}$Udaipur Solar Observatory, Physical Research Laboratory, Udaipur -313001, India
}

\date{Accepted XXX. Received YYY; in original form ZZZ}

\pubyear{2022}

\begin{document}
\label{firstpage}
\pagerange{\pageref{firstpage}--\pageref{lastpage}}
\maketitle

\setreviewsoff
\begin{abstract}
Although the enhancements in the alpha-proton ratio in the solar wind (expressed as $A_{He} = N_{a}/N_p*100$) in the Interplanetary Coronal Mass Ejections (ICMEs) have been studied in the past, $A_{He}$ enhancements at the stream interface region received very little attention so far.  In this letter, by extensively analyzing the stream interaction region (SIR) events observed in solar cycle 23 and 24, we show that the stream interface of alphas starts separating out from that of protons from the minimum of solar cycle 23.  \add{The population of alpha particles are enhanced compared to protons at higher angles between bulk velocity and local magnetic field (henceforth, bulk velocity angle) in the fast wind region of SIRs if the background solar wind is taken as reference.} \add{The analysis of} differential velocities \add{between} alphas and protons also reveals that the faster alpha particles accumulate near the fast wind side of the stream interface region leading to enhancement of $A_{He}$. The investigation brings out, for the first time, the salient changes in $A_{He}$ in SIRs for the two solar cycles and highlight the importances of \add{bulk velocity angle}  and differential velocity  in the fast wind region for the changes in $A_{He}$ in SIRs.

\end{abstract}




\section{Introduction}
The stream interaction regions (SIRs) are large-scale and long lasting structures in the Interplanetary (IP) medium. SIRs influence the dynamics of near-earth solar wind properties and hence, associated space weather impact. The SIRs are formed by the interaction of high-speed solar wind stream with the preceding slow solar wind. The increase in plasma density, magnetic field strength, and plasma pressure indicate compressed plasma in the stream interaction region.  SIRs have three parts - slow wind region, stream interface region, and fast wind region. The stream interface (SI) is usually characterized by an abrupt drop in density, an increase in the temperature/pressure, and a change in velocity with a large gradient at 1 au \citep{Burlaga1974, Belcher1971}. Earlier researchers have studied the changes in solar wind plasma parameters, i.e., density, magnetic field, dynamic pressure, etc. \citep[][and references therein]{Richardson2018}  associated with SIR.  However, there are very few studies \citep[e.g.][]{Gosling1978, Durovcova2019} on how stream interaction regions would alter the solar wind plasma composition in general and alpha-proton ratio (expressed as $A_{He} = N_{a}/N_p*100$) in particular. This is probably because large enhancement in $A_{He}$ in SIRs is rare unlike interplanetary Coronal Mass Ejections (ICMEs) wherein significant enhancements in $A_{He}$ are observed quite frequently \citep[][etc.]{Richardson2010, Fu2020, Yogesh2022}. Nevertheless, how $A_{He}$ varies across the stream interface can be very useful in characterising SIR structures and to evaluate their space weather impact.    

It is also worthwhile to mention here that $A_{He}$ can provide indication about the source region of the solar wind \citep[e.g.,][]{Borrini1982, Kasper2007}. It is also known that the although $A_{He}$ is 8\% in the photosphere, it gets depleted in corona and solar wind and gets fixed at a 4-5\% level \citep{Laming2004}. \cite{Yogesh2021} showed that this scenario got changed in solar cycle 24 when $A_{He}$ shifted towards lower values (2-3\%) indicating changes in the helium processing in the corona in the last cycle. Further, $A_{He}$ varies in solar wind according to solar wind speed and the solar activity level \citep{Kasper2007, Alterman2019, Yogesh2021}. Usually, the fast wind shows higher $A_{He}$ as compared to the slow wind. $A_{He}$ values are more variable in slow solar wind, whereas it does not vary significantly in the case of fast wind. \cite{Yogesh2022} also showed that the possible interplay of chromospheric evaporation and gravitational settling determines the enhanced $A_{He}$ level in ICMEs reported earlier \citep[e.g.][etc.]{Borrini1982, Fu2020}. However, very little has been reported for the $A_{He}$ variations in SIR. It was thought earlier that changes (generally increase) in helium abundance in SIR structures are only because of the transition in the type of solar wind \citep{Gosling1978, Wimmer1997}. \cite{Gosling1978} also showed that alpha flow speeds relative to protons change abruptly at the interface. Recently, \cite{Durovcova2019} suggested that the pitch angles and velocity of alphas in the proton frame are important factors to be accounted to explain the enhancement in $A_{He}$ in SIRs. 

	In this letter, we \add{investigate} a large number of SIR events spanning over solar cycle 23 and 24 (henceforth, SC23 and SC24) in the context of slow/fast wind and solar maxima/minima. This has not been attempted so far.  

\section{Data and selection of SIR events}
For the present investigation, we have used the Solar Wind Experiment (SWE) data on-board the WIND spacecraft  \citep{Ogilvie1995}. The dataset has a resolution of approximately 92s. The magnetic field measurements are taken from the Magnetic Field Investigation experiment \citep{Lepping1995}. 

The SIR events are taken from the catalogue compiled in \cite{Chi2018}. The details regarding the selection of events and the start and end times of the events at the spacecraft can be found in \cite{Chi2018} and references therein. This catalogue contains 866 SIR events observed during 1995 to 2016. We have removed the events which do not have continuous coverage of data. The SIR events, having possible mix-up with ICMEs within one day before the start time and one day after the end time, are also removed to make sure that only pure SIR events are considered. The ICME events are taken from the Richardson \& Cane catalogue \citep{Richardson2010}. The above criteria lead to 436 events that are used eventually for the present investigation. 

We have divided the SIR events into four categories, i.e., SC23 minima (1996-1998, 2006-2009), SC23 maxima (1999-2005), SC24 minima (2010-2011, 2016), and SC24 maxima (2012-2015). Note, SIR events beyond 2016 are not available in this catalogue making the number of SIR events considered in SC24 much less compared to SC23. Nevertheless, by considering identical number of SIR events, we have verified (see supplementary Figure S1 and compare with Figure \ref{fig:4}) that the scientific inferences reported in this work remain invariant. Therefore, in this work, we exploit the full available database.   

\begin{figure}
\begin{center}
\includegraphics[scale=0.70]{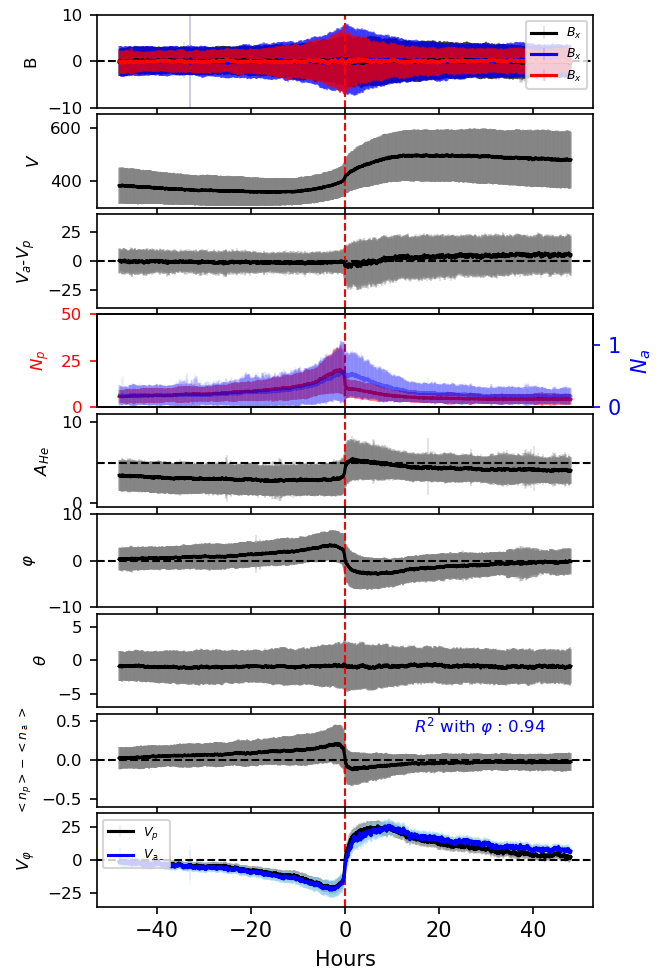}
\caption{Superposed epoch analysis (SPA) of the 436 SIR events. The red vertical dashed line  is the epoch time. The stream interface is chosen as the epoch time. The upper seven rows show the magnetic field components ($B_x$, $B_y$, $B_z$), velocity (V), the difference between the alpha and proton velocities ($V_a$ - $V_p$), number density of alpha and proton ($N_a$, $N_p$), Helium abundance ($A_{He}$), East-West GSE bulk Flow Angle ($\phi$) and North-South flow angle ($\theta$). The  row number eight shows the difference between scaled  proton number density (Scaled density = <N> =(n-$n_{min}$)/($n_{max}$-$n_{min}$)) and scaled alpha number density (<$N_p$> - <$N_a$>). The   row number nine shows the East-West velocity of alphas and protons ($V_{\phi}(a)$, $V_{\phi}(p)$). The $\phi$ and $V_{\phi}$ are good indicators for the change in the solar wind type, i.e., from slow to the fast wind. Note, $R^2$ (coefficient of determination) between <$N_p$> - <$N_a$> and ${\phi}$ is very high (more than 90\%). This suggests that the proton pile-up dominates slow wind, whereas the alpha pile-up dominates fast wind.       \label{fig:1}}

\end{center}
\end{figure}
 	
\section{Results and Discussion }
\subsection{Superposed Epoch Analysis (SPA) of the total events during solar maxima and minima }

Figure \ref{fig:1} shows the superposed epoch analysis (SPA) of the SIR events used in this work. The cadence of data is changed from the 92s to 2 minutes to generate the SPA outputs.  The stream interface (SI) is the boundary between the slow and fast wind. The SPA is carried out with respect to the zero epoch line which, in this case, is the SI. Henceforth, the slow wind side of SI will be referred to as the Slow Wind Region (SWR), and the fast wind side will be described as the Fast Wind Region (FWR). It can be noted from Figure \ref{fig:1} that although SPA mixes together measurements from inside the SIRs with those from outside the SIRs, the characteristic changes related to SI occur mostly within an interval of ±10 hours with respect to zero epoch line.  Note, in our investigation, the duration of 80\% of SIR events are more than 20 hours. Therefore, we believe that results presented in this work are statistically valid.

The change in the magnetic field, number density, and velocity can be observed in the SPA (Figure \ref{fig:1}). The change in East-West flow angle ($\phi$) is a good indicator for SI identification \citep{Mayank2022, Rout2017}. An important point that emerges from this Figure is that the difference between the scaled proton number density (Scaled density = <N> =(n - $n_{min}$)/($n_{max}$ - $n_{min}$)) and scaled alpha density show very high correlation (See Figure \ref{fig:1}) with the East-West flow angle ($\phi$). This differential scaled density also suggests that the pile-up of protons is dominant over alphas in SWR, whereas the alphas pile-up dominates towards FWR. Therefore, it becomes apparent from the SPA that this differential pile-up is the primary cause of relative enhancement of $A_{He}$ in the FWR of SIRs. The possible reasons for this differential pile-up and the variations in $A_{He}$ in minima and maxima of SC23 and SC24 are taken up in the ensuing sections. 

\begin{figure}
\begin{center}
\includegraphics[scale=0.18]{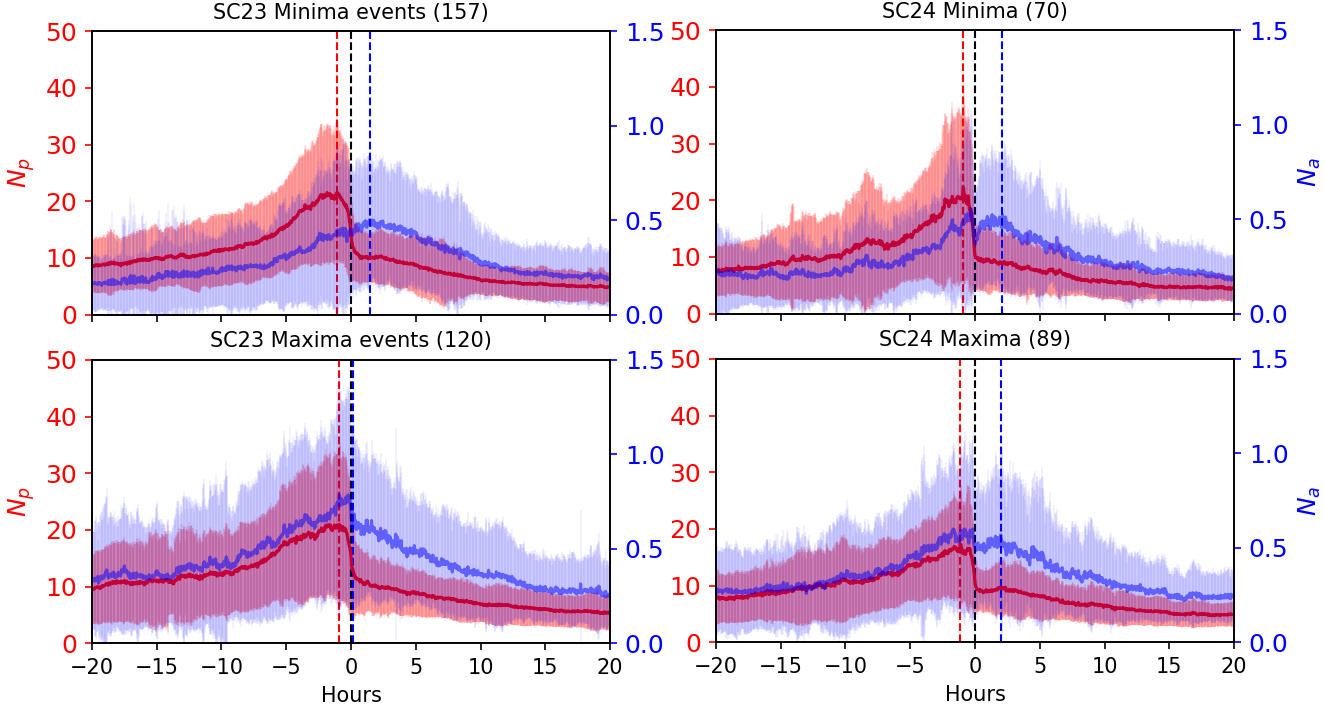}
\caption{SPA of number densities of alphas (blue) and protons (red) in SC23 and SC24 are shown with one sigma error bar for SC23 minima, SC24 maxima, SC24 minima, and SC24 maxima. The number of events considered are also mentioned at the top of each panel. The dashed black, red, and blue vertical lines represent the stream interface (SI), the peak value of proton density towards the SWR, and the peak value of alpha density towards the FWR, respectively. The additional peaks of alpha particles towards fast wind (in SC23 minimum and SC24) suggest the differential behavior between the alphas and protons across SI.     \label{fig:2}}

\end{center}
\end{figure}

\subsection{Superposed Epoch Analysis (SPA) of number densities of alphas and protons in SC23 and SC24}

In this section, we show the SPA of the number density of alphas and protons ($N_a$ and $N_p$) corresponding to maxima and minima of SC23 and SC24 (Figure \ref{fig:2}).  The stream interface (dashed black line) is used as the zero-epoch time. The duration of 20 hours before and 20 hours after the zero-epoch time is shown. The variations in protons and alphas are shown in red and blue color respectively, with one sigma error bar. The red and blue dashed vertical lines represent the peak value of proton number density towards SWR and the peak alpha number density towards FWR.

It can be observed from Figure \ref{fig:2} that alphas have distinct additional SIs towards FWR during SC24 (both minima and maxima) and this additional SI seems to be absent in SC23. In SC23, the SI of alphas either coincides with SI of protons (maxima) or lies distinctly separated (primary SI) in the FWR region (minima). Whether it is primary or additional alpha SI, both are separated  by $\sim$ 2 hours from the proton SI. Therefore, as the alpha SIs are well separated  (with the exception of SC23 maxima) based on statistically significant dataset, we treat these alpha SIs as real. In addition, there is no abrupt decline in the alpha number density similar to proton density across the SI. This suggests that an enhanced helium abundance can be expected in the FWR of SIRs. It is now known that SC24 is a weaker cycle and the declining solar activity could be observed from the SC23 minima (deep minimum) itself while the maximum of SC23 was relatively stronger \citep{Hathaway2015}. Therefore, it appears that the enhanced helium abundance at the SWR of SC23 maxima is associated with the SC23 solar activity.

\cite{Durovcova2019} considered the SIRs as analogous to a magnetic mirror assembly. The important parameters which control the motion of charged particles in curved magnetic fields are the pitch angle and the velocity of ions. \add{In this work, instead of pitch angle, we have considered angle between bulk velocity vector and local magnetic field vector and differential velocity between alphas and protons.} We have evaluated how these parameters affect the alpha and proton number density for different phases of SC23 and SC24. It can be noted from Figure \ref{fig:2} that the behavior of alphas and protons are different in the SWR and FWR of SIR. Therefore, these regions are investigated individually in the upcoming subsections.

\add{\subsection{Bulk velocity Angle Distribution (BAD)}}
We define \add{bulk velocity angle with respect to the local magnetic field (in short, BA) } for both protons and alphas and \add{construct Bulk velocity Angle Distribution (BAD).} \add{Note, BAD is different from pitch angle distribution or PAD as BAD deals with the angle of bulk velocity of ions with respect to the local magnetic field direction while PAD deals with the  angles between the individual ion velocity with respect to the local magnetic field direction. BA is defined as follows} 

\begin{equation}
BA= cos^{-1}(\frac{\vec{B}.\vec{V}}{|\hat{B}||\hat{V}|})
\end{equation}

Here B is the magnetic field vector and V is the bulk velocity vector. The  \add{BAD} from $0^o$ to $90^o$ is only considered here and this is ensured by taking modulus in the denominator. This is done as the \add{BAD} between $90^o$ to $180^o$ is expected to be the mirror image of the  \add{BAD} between $0^o$ to $90^o$. To start with, we have grouped the \add{BA} data ($\sim$ 92 sec cadence) from SWR and FWR of all the SIR events separately. Subsequently, a \add{BAD} is constructed by taking a bin size of $2^o$.  Figure \ref{fig:3} shows the \add{BAD} in minima and maxima of SC23 and SC24 in SWR. To understand the effect of the stream interaction, the  \add{BAD} of SIRs is compared with the \add{BAD} of the background slow and fast winds in absence of SIR. Solar wind velocities less than 400 km/s and higher than 500 km/s are used to construct the references for the slow and fast background wind respectively. The solar wind having velocities between 400-500 km/s is not considered to maintain the sanctity of the references. In addition, the ICME events are also removed from the data to estimate the background references. Note the references are constructed based on data during 1996-2016 to make it consistent with the SIR catalog.  The red and blue colors are for protons and alphas respectively.

\subsubsection{Slow Wind Region (SWR)}

\begin{figure}
\begin{center}
\includegraphics[scale=0.42]{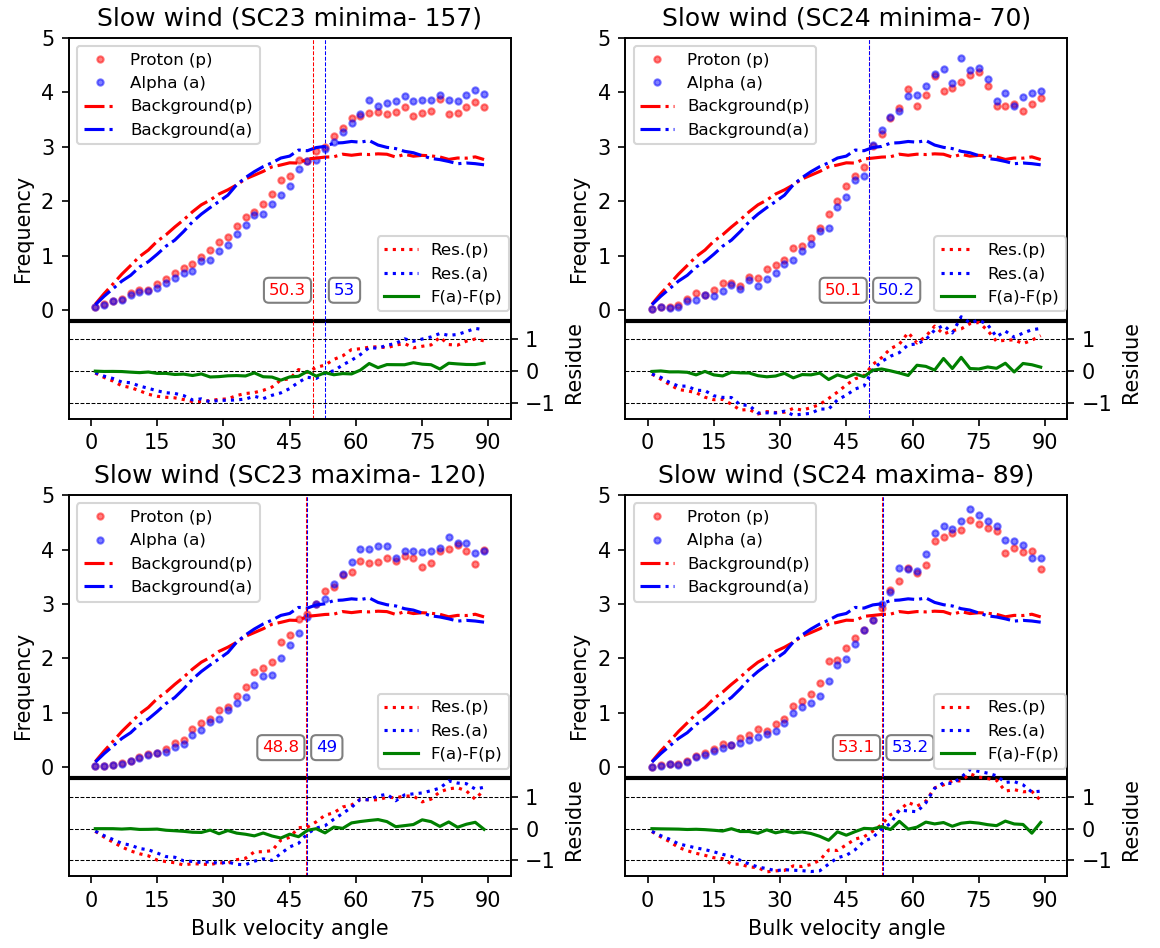}
\caption{The  \add{BAD} for minima and maxima for SC23 and 24 are shown only for the slow wind region (SWR). The red and blue colors are used for protons and alphas, respectively. The upper part of each panel shows the \add{BAD} for SWR of SIR and slow background wind. The lower part of each panel shows the difference between SIR \add{BAD} and the background slow wind. The green color indicates the difference between alpha and proton \add{BAD} for SWR.      \label{fig:3}}

\end{center}
\end{figure}

In each panel of Figure \ref{fig:3}, the upper part shows the \add{BAD} (of protons and alphas) in SWR of SIR in dotted lines whereas the dotted dashed lines capture the \add{BAD} for the background wind. The dashed vertical lines show the cross-over of the SWR of SIR and the background \add{BAD}. The lower part of each panel shows the difference between the \add{BAD} in SWR of SIR and background slow solar wind. The green line shows the difference or residual between the \add{BAD} of alphas and protons in the SWR of SIR. 

A few important points can be noted from Figure \ref{fig:3}. A part of the lower \add{BAs} (before the crossover) distribution seems to have shifted towards the higher values (after the crossover). This shift seems to maximize in the maxima of the SC24. The crossover point is nearly at $50^o$ in SC23 and SC24. The alphas and protons have similar crossover points except for SC23 minima wherein a difference of $2.7^o$ is noticed. The SC24 maxima shows the highest shift of particles from lower \add{BAs}  to higher \add{BAs}. Another important feature is that the SWR distribution shows a plateau region (lack of distinctly sharp peaks in the \add{BAD}) for both maxima and minima for higher \add{BAs}  in SC23. In contrast, conspicuous peaks are noticed in the case of SWR distribution during both maxima and minima in SC24. Despite all the above features, the green line is near zero which means there is little difference between the alpha and proton \add{BAD}. Therefore, there is no significant difference between the \add{BAD} of alphas and protons in the SWR of SIR although there are higher number of protons or alpha particles for higher \add{BAs} than the background slow solar wind.

\subsubsection{Fast Wind Region (FWR)}

\begin{figure}
\begin{center}
\includegraphics[scale=0.42]{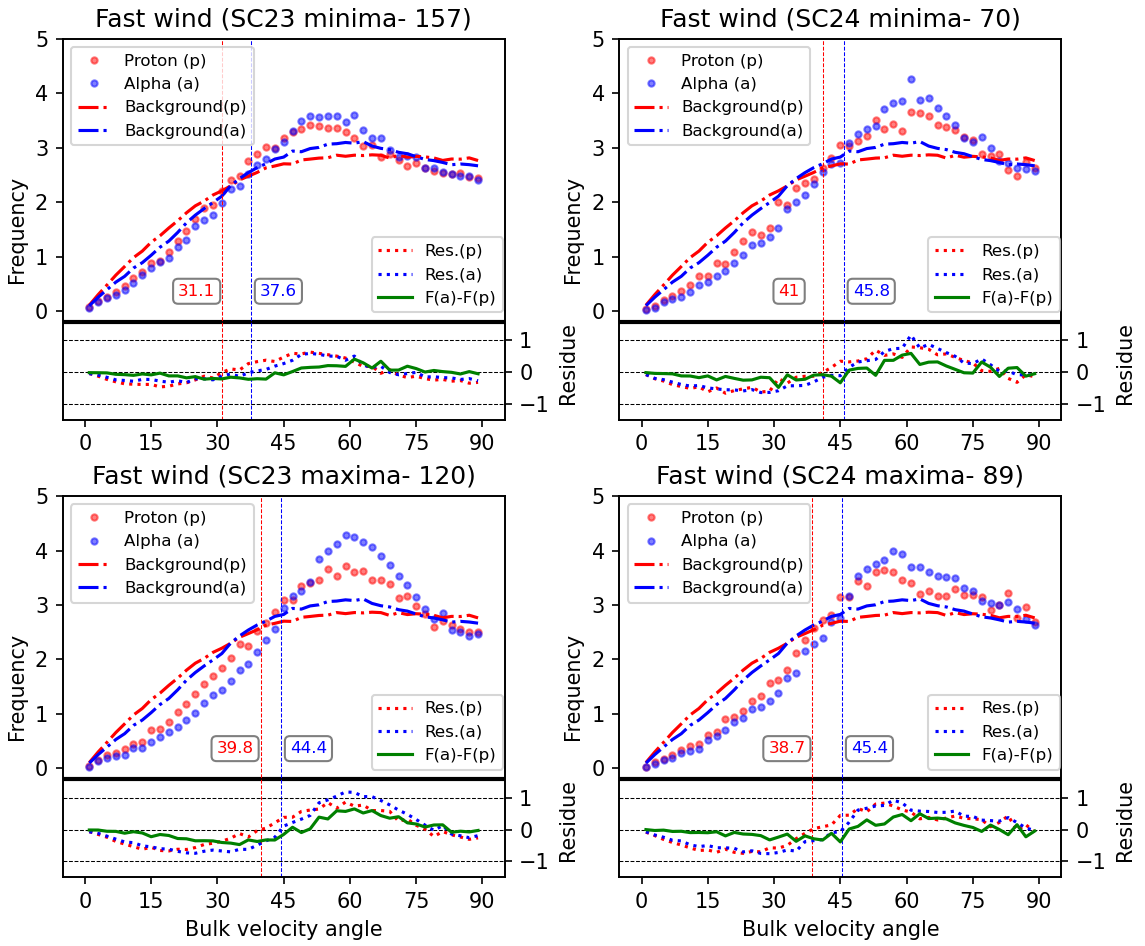}
\caption{The \add{BAD} for minima and maxima for SC23 and 24 are shown only for the fast wind region (FWR). The red and blue colors are used for protons and alphas, respectively. The upper part of each panel shows the \add{BAD} for FWR of SIR and background fast wind. The lower part of each panel shows the difference between SIR \add{BAD} and background fast wind. The green color indicates the difference between alpha and proton \add{BAD} for FWR.      \label{fig:4}}

\end{center}
\end{figure}

A similar analysis is also performed for FWR in SIR. Figure \ref{fig:4} is analogous to Figure \ref{fig:3} but with the properties of FWR and background fast wind. It can be observed from Figure \ref{fig:4} that a part of particle distribution having lower \add{BAs} than the background solar wind seems to have shifted towards higher values similar to SWR. However, one difference is noticed. The \add{BAD} for FWR and background fast wind have at least two distinct crossover points unlike the slow solar wind when only one crossover point is seen. The first crossover point is at the lower \add{BA}, whereas the second is at a very high value. The crossover \add{BAs} are nearly $40^o$ and $45^o$ for protons except for SC23 minima ($31.1^o$ (proton) and $37.6^o$ (alpha)). The proton and alpha \add{BAs} crossover points have a difference of approximately $5^o$. The alphas have higher crossover angles, suggesting that more alpha particles are shifted towards higher \add{BAs}. Unlike SC23, the variation in the green line (in the lower panels) in this case suggests that alphas are more at higher \add{BAs} than protons. This alpha and proton distribution difference is the least in the SC23 minima. Note, SC23 is characterized by a deep minimum period \citep{Hathaway2015}.  Based on these results, it can be stated that the alphas behave differently than protons in the case of the FWR. In addition, the \add{BA} crossovers are different from those for SWR.  	

We have also repeated this exercise for helium ions in solar wind (proton) frame. These results are presented in supplementary Figures S2 and S3. It can be seen from S2 and S3 that the alpha particles are shifted towards higher \add{BAs} in FWR whereas this feature is absent for SWR. This is consistent with Figures \ref{fig:3} and \ref{fig:4}.  

\subsection{Distribution of differential velocity between alphas and protons in SC23 and 24 }

\begin{figure}
\begin{center}
\includegraphics[scale=0.42]{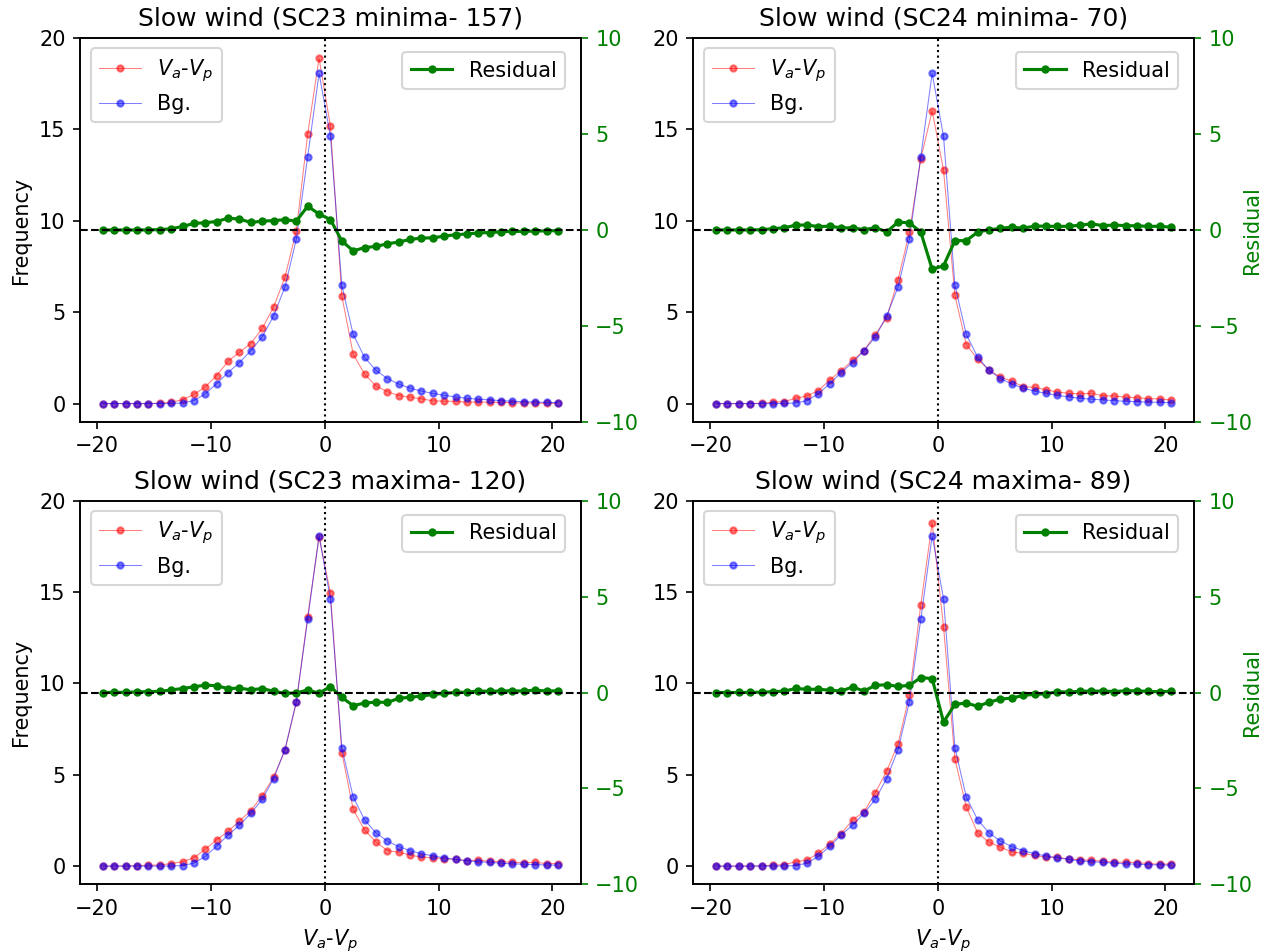}
\caption{The distribution of differential velocity (alpha velocity – proton velocity) for minima and maxima for SC23 and 24 are shown only for the slow wind region (SWR). The red and blue colors are used for differential velocity distribution for the SWR in SIR and background (Bg.) slow wind respectively. The green color shows the difference between the slow wind region in SIR and the background slow wind distribution..      \label{fig:5}}

\end{center}
\end{figure}

\begin{figure}
\begin{center}
\includegraphics[scale=0.42]{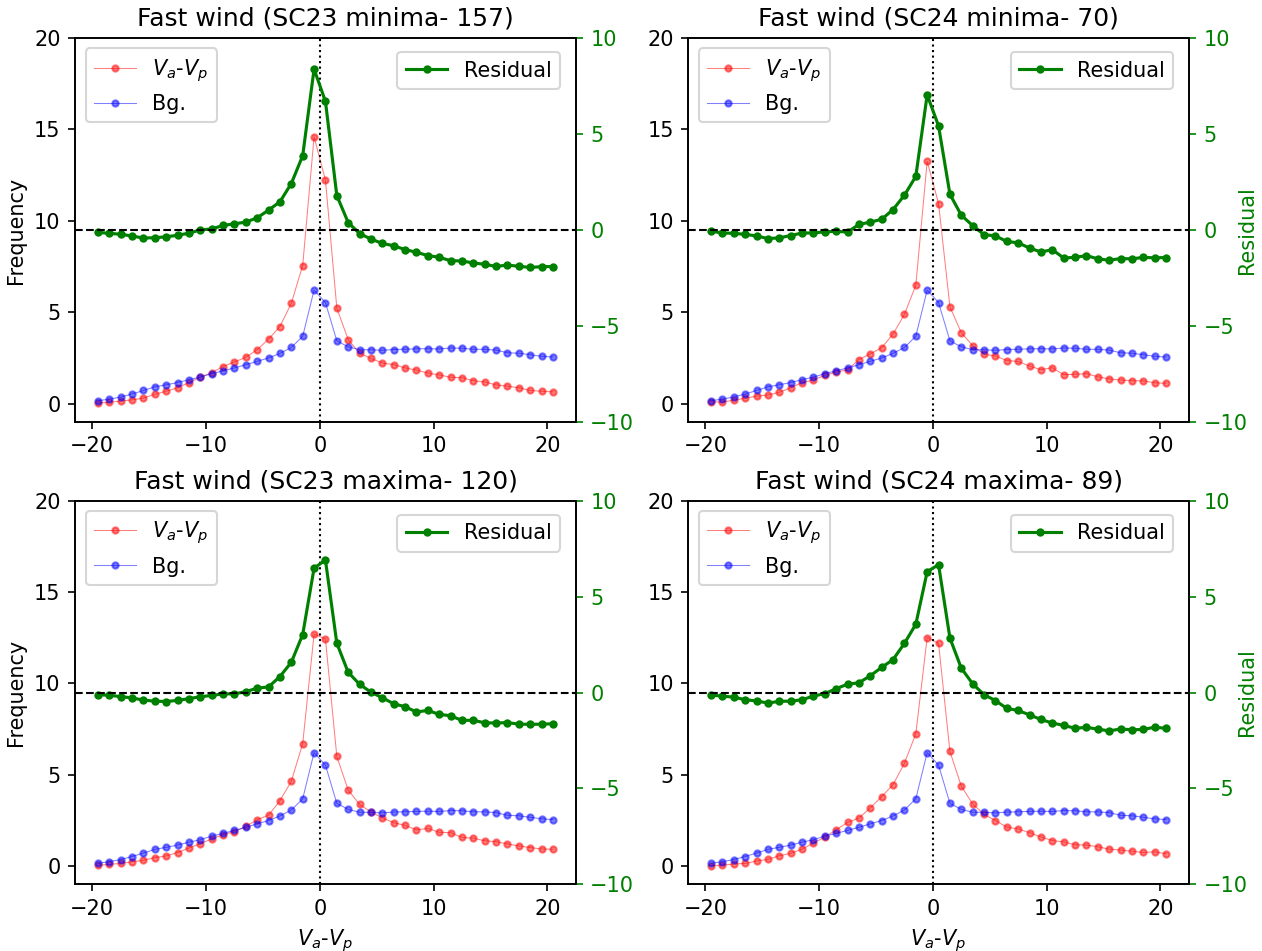}
\caption{The distribution of differential velocity (alpha velocity – proton velocity) for minima and maxima for SC23 and 24 are shown only for the fast wind region (FWR). The red and blue colors are used for differential velocity distribution for the FWR of SIR and background (Bg.) fast wind respectively. The green color shows the difference between the fast wind region of SIR and background fast wind distribution.      \label{fig:6}}

\end{center}
\end{figure}

The variation in the difference between the alpha and proton velocities (Differential velocity) are shown in Figures \ref{fig:5} and \ref{fig:6} for SWR and FWR in SC23 and 24 similar to Figures \ref{fig:3} and \ref{fig:4}. Figures \ref{fig:5} and \ref{fig:6} show the distribution of velocity difference for SWR and FWR respectively. The red color in both the Figures shows the differential velocity distribution for SWR and FWR for SIRs. The blue color indicates the distribution for the background differential velocity. The green color is the difference between the two distributions, i.e., differential velocity distribution for SIR and background velocity. It can be observed from Figure \ref{fig:5} that there is hardly any difference in the distribution for SWR in SIR and in the background solar wind barring slight differences at the distribution center ($V_a$ - $V_p$ = 0). 

The FWR distribution for differential velocity shows significant changes than the SWR distribution. The background distribution shows relatively enhanced tails towards the positive side because of the fact that alphas have higher velocity than protons in the background fast wind. Or in other words, velocities of alpha particles get reduced in the case of the FWR of SIRs. Therefore, the faster alpha particles are slowed down in the interaction region and accumulate near the center (Va=Vp) of the distribution. This is captured by the substantial enhancement of the residual curve in green near the center, suggesting the accumulation of alpha particles in the proton frame. This is the primary cause of the enhanced helium towards the FWR of SIRs  and it is probably consistent with additional SI found in Figure \ref{fig:2}. The additional SI for alphas in Figure \ref{fig:2} suggests preferential accumulation of alpha particles in the FWR. The decoupling of this additional SI from the SI determined based on proton number density, compressed magnetic field at the interface, flow deflection, and temperature increase \citep{Chi2018} suggest that these additional alpha SI is not because of mixing. If it were the case, additional SI would have been found in SWR also. Therefore, it appears that alpha particles get decoupled towards the FWR owing to differential forcing arising out of the differences in \add{BA} and velocity with respect to protons. 

We also note that there is little difference in the behaviours of velocity differences in the maxima and minima of SC23 and SC24. This suggests that unlike $N_{a}$ and \add{BA}, the change in the differential velocities between alphas and protons in SIRs with respect to the background solar wind is due to primarily interplanetary processes. This aspect will be discussed further in the next section.   

\subsection{Validity of magnetic mirror hypothesis }

It must be mentioned that processes other than magnetic mirror may also contribute in the variation of $A_{He}$ within SIRs. As the SPA shows distinctive changes in $A_{He}$ around SI, the role of other processes operational within the interaction region cannot be neglected. Therefore, the finite role of diffusive shock acceleration in changing $A_{He}$ in SIR may be hypothesized \citep{Durovcova2019}. However, as \cite{Durovcova2019} pointed out, changes in $A_{He}$ are also observed for SIRs/CIRs that are not bounded by shocks. Under this scenario, the role of diffusive shock acceleration is not unambiguous. Further, the simple magnetic mirror model disregards the effects of turbulence and plasma waves in the solar wind that may be associated with the passage of SIR through IP medium. Note, propagation of SIR/CIR through IP medium can bring in non-linear effects associated with magnetic field steepening and expansion \citep[e.g.][]{Burlaga2004,Durovcova2019}. For an accurate description and modelling of the behaviour of helium inside SIRs, these aspects need to be considered. 

\subsection{Uncertainties associated with alpha particle measurements}
 The derived alpha parameters, in general, come with larger uncertainties than proton parameters. In addition, in the compression region, there can be significant overlap between proton and alpha Velocity Distribution Functions (VDFs). In the present work, we have considered the mean values of proton, alpha densities and velocities without considering the 1-sigma (standard deviation) variations. This will not affect the inferences drawn in this paper although the exact nature of \add{BA} and differential velocity distributions will be affected. For example, if we consider 1-sigma variations, the value of the crossover \add{BA} will change as well as the width of the residual curve centred around zero differential velocity will get affected.

\section{Summary}
 In this letter, we show that alphas and protons show different behavior in the SIRs and enhancement in $A_{He}$ occurs across the stream interface in the fast wind region. \cite{Gosling1978} suggested that the enhancement in $A_{He}$ at the stream interface is caused by the difference in the type (fast or slow wind having high and low $A_{He}$ respectively) of solar wind. The SIRs can be considered as magnetic bottles or mirror assemblies \citep{Durovcova2019}. It is well known that the magnetic field and the velocity of the ions play important roles in controlling the bounce motion of particles. The magnetic mirror force depends on the ions' magnetic moment, m = m$V_{\perp}{^2}$/2B. This force slows down the ions or the ions having a \add{BA} higher than the loss cone angle tend to have bounce-back motions. Our results suggest that alphas and protons have similar \add{BAs} in SWR, but in the case of FWR, more alpha particles are distributed towards higher \add{BAs} as compared to protons. It may be noted that the mirroring hypothesis works best when \add{BAs} (varying 0 to $90^o$ in this case) exceed the loss cone angle. Therefore, as long as the mean loss cone angle is closer to the \add{BA}  crossover shown in Figures \ref{fig:3} and \ref{fig:4}, the mirroring hypothesis can be applied. The observed changes in $A_{He}$ in SIR with respect to loss cone angle is discussed in detail in \cite{Durovcova2019}. Another aspect to be noted is that the \add{BA} crossover between SIR distribution and background distribution is similar for alpha and protons for SWR. This crossover differs by approximately $5^o$ in the case of FWR of SIRs. This change in distribution may play an important role in causing the enhancement of alpha particles in the FWR of SIR. 
 
In addition to \add{BA}, solar wind velocity also plays a vital role in deciding the number density. In general, protons and alphas have similar speeds in the slow wind, whereas alphas are faster than protons in the fast wind. The alphas have a higher magnetic moment because of the higher velocity and mass. So, alpha particles experience more magnetic curvature or mirror force than protons. The implication of this higher force can be seen in Figure \ref{fig:6} as the frequencies of alpha particles faster than protons are reduced in the case of FWR of SIRs. So, this difference in curvature force for alphas and proton causes a generation of the second peak of alpha particles towards FWR. 

  	The additional important point that emerges from this work is that the SWR \add{BAD} in SC24 shows similarities with the overall FWR  \add{BAD} as these show peaks    at higher \add{BAs}. On the contrary, the SWR distribution in SC23 shows a plateau region at higher \add{BAs}. This probably indicates additional influence of the changes in the sources of slow solar wind in SC24 compared to SC23. 

The above results reveal that the variations in helium abundance near the stream interface (SI) are not primarily determined by the level of solar activity. Instead, it appears that the primary factors affecting helium abundance in this region are \add{BA} and differential velocity. The solar activity can determine the level of enhancement towards SWR because it changes the background value of helium abundance in the slow solar wind. This can be observed in Figure \ref{fig:2} that the level of enhancement in helium differs in SWR whereas these levels are approximately the same in the FWR region.

In a nutshell, the SWR of SIRs does not show significant changes in \add{BA} and differential velocity. This may be the primary cause behind the similar enhancement in the alphas and protons  in the case of SWR. In the case of FWR, the \add{BA} and differential velocity distribution show significant differences between alphas and protons. The crossover between background distribution and the FWR distribution for alphas is at a higher angle than protons. It means that more alphas are shifted towards higher \add{BA} than protons, causing a significant enhancement in the alpha number density. This probably causes the second peak of alphas in the FWR. The differences in the \add{BAD} for SC23 (plateau) and SC24 (peak) possibly indicate additional changes in the source of slow wind during SC23 and SC24. On the contrary, the differential velocities of alphas and protons do not show any significant change in SC23 and SC24. This suggest primary role of SIR in determining the differential velocities.

\section*{Acknowledgements}

We thank the SWE (WIND) team for their efforts. We acknowledge Coordinated Data Analysis Web for its open data policy. This work is supported by the Department of Space, Government of India.

\section*{Data Availability}
The data can be obtained from \url{https://cdaweb.gsfc.nasa.gov/index.html/}.



\bibliographystyle{mnras}
\bibliography{ref} 



\label{lastpage}

\end{document}